\documentclass[amsmath,amssymb,nofootinbib,superscriptaddress,showkeys,twocolumn]{revtex4-2} 
\usepackage{graphicx} 
\usepackage{dcolumn} 
\usepackage{bm} 
\usepackage{subfigure} 
 \usepackage{multirow} 
\usepackage[colorlinks=true, citecolor=blue, urlcolor = magenta, linkcolor= red, bookmarks=true]{hyperref} 
\usepackage{orcidlink} 
\usepackage{booktabs} 
\usepackage{csquotes}
\graphicspath{{figs/}} 

\begin{document}

\title{Periodic orbits as probes of charged loop quantum gravity black holes through gravitational waves} 

\author{Abolhassan Mohammadi \texorpdfstring{\href{https://orcid.org/0000-0003-1228-9107}{\orcidlink{0000-0003-1228-9107}}{}}}\email{abolhassanm@hnit.edu.cn; @gmail.com}
\affiliation{School of Science, Hunan Institute of Technology, Hengyang 421002, China.}
\author{Arun Kumar\texorpdfstring{\href{https://orcid.org/0000-0001-8461-5368}{\orcidlink{0000-0001-8461-5368}}{}}}\email{arunbidhan@gmail.com}
\affiliation{Centre for Theoretical Physics, 
	Jamia Millia Islamia, New Delhi 110025, India}

\author{Hongwei Tan 
}
\email{honweitan@hnit.edu.cn}
\affiliation{School of Science, Hunan Institute of Technology, Hengyang 421002, China.}

\author{Sushant G. Ghosh \texorpdfstring{\href{https://orcid.org/0000-0002-0835-3690}{\orcidlink{0000-0002-0835-3690}}{}}}\email{sghosh2@jmi.ac.in}
\affiliation{Centre for Theoretical Physics, 
	Jamia Millia Islamia, New Delhi 110025, India}
\affiliation{Astrophysics and Cosmology Research Unit, 
	School of Mathematics, Statistics and Computer Science, University of KwaZulu-Natal, Private Bag 54001, Durban 4000, South Africa}

\begin{abstract}
Gravitational waves from extreme--mass--ratio inspirals (EMRI) provide a direct probe of the strong--field geometry of black holes. Motivated by this, we study the motion of test particles and the resulting gravitational wave emission in the spacetime of a charged black hole inspired by loop quantum gravity (LQG), where the classical singularity is replaced by a smooth transition surface arising from the LQG polymerization, in which its radius is set by the LQG area gap condition. As a result, the polymerization parameter $\delta_b$ is uniquely determined by the mass $M$ and charge parameter $Q$, so that all cases examined in this work contain LQG correction. By constructing the effective potential, the innermost stable circular orbit (ISCO) and the marginally bound orbit (MBO) are determined. Periodic orbits are classified using the Levin--Perez--Giz zoom--whirl taxonomy,  showing how the orbit topology shapes the waveform, so that each closed trajectory is labeled by the triple integer $(z, w, v)$ and located through the rational frequency ratio $q = \omega_\phi/\omega_r - 1$. Within the quadrupole approximation, the gravitational waveforms for an EMRIs are estimated, and the resulting polarizations are obtained in the time--domain and frequency--domain. The resulting polarizations in the time--domain exhibit a zoom--whirl morphology, with the waveform amplitude and phase dependent on the LQG parameter. The characteristic strain peaks in the millihertz band for all values of the charge parameter $Q$, and they exceed the projected sensitivities of LISA, Taiji, and TianQin, suggesting that future observations could place meaningful constraints on the LQG polymerization parameter in the strong-field regime. 
\end{abstract}
	
	
\maketitle

\section{Introduction}\label{sec1}
General relativity (GR) has passed every experimental test we have thrown at it, from solar system dynamics to gravitational waves from binary black holes~\cite{LIGOScientific:2016aoc, LIGOScientific:2016vbw, LIGOScientific:2016vlm, LIGOScientific:2016emj}. But we know it cannot be the whole story. The singularities at the centers of black holes signal a breakdown of classical gravity, and the theory refuses to play nicely in the quantum regime. The singularity problem in classical black holes motivates the search for quantum--gravity completions of black holes or the development of a quantum theory of gravity.

Loop quantum gravity (LQG) is one of the leading candidates for a quantum theory of gravity~\cite{Ashtekar:2004vs, Thiemann:2008zza, Rovelli:2015gwa}. Instead of trying to quantize gravity on a fixed background, LQG quantizes geometry itself. The result is a discrete spacetime at the Planck scale, with a smallest possible area, the area gap~\cite{Modesto:2008im}. When you apply LQG ideas to black holes, the singularity often disappears, replaced by a smooth bounce or a transition surface~\cite{Ashtekar:2018lag, Ashtekar:2020ckv}. These so-called LQG black holes are regular, and they make concrete predictions about the strong-field regime.

The implementation of LQG ideas in black hole spacetimes proceeds through polymerization, in which classical phase--space variables are replaced by bounded trigonometric functions of the connection. This modifies the interior Hamiltonian so that the evolution does not terminate at a singularity but instead continues through a bounce surface, the transition hypersurface, at which the spacetime undergoes a topological change from a black hole interior to a white hole interior~\cite{Modesto:2010uh, Corichi:2015xia, Ashtekar:2018cay, Ashtekar:2018lag}. Several polymerization schemes have been explored in the literature~\cite{Bodendorfer:2019jay, Bodendorfer:2019nvy, ElizagaNavascues:2022rof, ElizagaNavascues:2022npm, Alonso-Bardaji:2022ear, Alonso-Bardaji:2021yls, Moreira:2023cxy}, with the model of Alonso-Bardaji, Brizuela, and Vera being particularly notable for producing a symmetric black--to--white hole transition in which both phases share the same mass and the exterior metric is asymptotically flat~\cite{Alonso-Bardaji:2022ear, Alonso-Bardaji:2021yls}.

The catch is that quantum gravity effects are usually tiny, suppressed by powers of the Planck length. To see them, you need a system that magnifies small deviations, and black holes do that. Extreme mass-ratio inspirals (EMRI), where a stellar-mass compact object spirals into a supermassive black hole, are particularly good at this~\cite{LISA:2022yao, Babak:2017tow}. The small body completes thousands or even millions of orbits in the strong field, and any tiny deviation from GR can accumulate into a measurable phase shift in the gravitational waveform~\cite{Barack:2003fp, Gair:2012nm}.

In this paper, we ask whether a specific LQG-inspired black hole leaves a detectable imprint on EMRI gravitational waves. The model we use comes from Borges et al.~\cite{Borges:2023fub}, who applied a polymerization scheme to the Reissner-Nordstr\"{o}m Hamiltonian and imposed the LQG minimal area condition. The result is a charged black hole with a smooth transition surface instead of a singularity. The polymerization parameter $\delta_b$ is not free; it is fixed by the mass $M$ and charge $Q$ through the area gap condition. So when we vary $Q$, we are also varying the strength of the LQG corrections.

Periodic orbits furnish a powerful tool for analyzing EMRI dynamics. Levin and Perez--Giz ~\cite{Levin:2008mq, Levin:2009sk} introduced a systematic taxonomy that classifies periodic orbits by three topological integers $(z,w,v)$, representing zoom, whirl, and vertex numbers. This classification scheme has been applied across a wide range of spacetimes, including Schwarzschild and Kerr black holes~\cite{Levin:2008ci, Levin:2009sk, Bambhaniya:2020zno, Rana:2019bsn}, charged black holes~\cite{Misra:2010pu}, hairy black holes in Horndeski theory~\cite{Lin:2023rmo}, Kerr--Sen spacetimes~\cite{Liu:2018vea}, naked singularities~\cite{Babar:2017gsg}, and a broad range of regular and modified--gravity geometries~\cite{Yao:2023ziq, Lin:2022llz,Gong:2021jgg, Chan:2025ocy, Wang:2022tfo, Haroon:2025rzx, Tu:2023xab, Zhou:2020zys, Alloqulov:2025bxh, Wei:2025qlh, Wang:2025wob}. The associated gravitational waveforms and characteristic strain spectra have been computed to explore quantum--gravity and modified--gravity corrections measurable imprints on the emitted radiation~\cite{Tu:2023xab, Yang:2024lmj, Zhao:2024exh, Jiang:2024cpe, Haroon:2025rzx, Alloqulov:2025ucf, Zare:2025aek, Gong:2025mne, Li:2025sfe, Deng:2025wzz,Kumar:2026hfx}.

We first analyze timelike geodesics in this spacetime and compute the innermost stable circular orbit. (ISCO) and the marginally bound orbit (MBO), and classify periodic orbits using the zoom--whirl taxonomy. We then generate the gravitational waveforms using the numerical kludge method, which combines exact geodesic integration with the quadrupole formula, following the approach of Babak \emph{et al.}~\cite{Babak:2006uv, Gair:2005ih}. This method has been widely used to study EMRI waveforms in modified--gravity theories~\cite{Qiao:2024gfb, Alloqulov:2025bxh, Yang:2025esa}. Finally, we compute the characteristic strain and compare the signals to the sensitivity curves of LISA, Taiji, and TianQin to see whether they could actually be detected~\cite{LISA:2017pwj, Hu:2017mde, TianQin:2015yph}.

What we find is that LQG corrections leave clear marks on the waveforms. The amplitude of the whirl--phase bursts grows with $Q$, and the waveform accumulates a phase shift that also increases with $Q$. Both effects show up in the frequency--domain as a higher, slightly blue--shifted characteristic strain peak. The signals fall squarely in the millihertz band and exceed the sensitivity thresholds of all three detectors. So if these black holes exist, future space--based observatories have a real chance of seeing them.

The paper is organized as follows. In Sec.~\ref{sec:lqgbh}, we review the charged LQG black hole metric and its key properties. Section~\ref{sec:geodesics_orbits} analyzes timelike geodesics, the ISCO, the MBO, and periodic orbits. Section~\ref{sec:gw} presents the gravitational waveforms, the frequency-domain analysis, and the characteristic strain. We conclude in Sec.~\ref{sec:conclusion} with a discussion of what we have learned and what comes next.
We use geometric units with $G = c = 1$ unless stated otherwise.

\section{LQG and Charges BH}\label{sec:lqgbh}

LQG is one of the leading approaches to quantum gravity, providing a non--perturbative and background--independent quantization of spacetime geometry~\cite{Ashtekar:2004vs, Thiemann:2008zza, Rovelli:2015gwa}. A key prediction of the theory is that geometric quantities such as area and volume acquire discrete spectra. The smallest non--zero eigenvalue of the area operator,
$\Delta = 4\pi\sqrt{3}\,\gamma\,\ell_{\rm Pl}^{2}$,
introduces a fundamental length scale into the quantum description of spacetime~\cite{Immirzi:1996dr, Meissner:2004ju, Ghosh:2004rq}. While these effects are negligible in weak--field regimes, they are expected to become relevant in regions of large curvature, such as black holes, where classical general relativity breaks down~\cite{Bojowald:2005epg, Modesto:2010uh, Ashtekar:2018lag}. In effective LQG models, the resulting quantum corrections are encoded through polymerization, which modifies the classical dynamics and replaces the spacetime singularity with a regular quantum--gravity core~\cite{Modesto:2010uh, Ashtekar:2018lag}.

In the polymerization programme, classical phase--space variables are replaced by bounded, trigonometric functions of the connection, so that the Hamiltonian acquires quantum corrections that become significant only at Planck--scale curvatures. A particularly elegant single--parameter scheme, developed by Alonso--Bardaji \emph{et~al.}~\cite{Alonso-Bardaji:2022ear, Alonso-Bardaji:2021yls}, yields a \emph{symmetric} black hole--to--white hole transition, wherein the two phases share the same mass, the exterior spacetime is asymptotically flat, and the transition surface lies inside the event horizon for any value of the polymerization parameter. Building upon this framework, Borges \emph{et al.}~\cite{Borges:2023fub} further expanded the construction to charged, spherically symmetric black holes. Starting from the classical Reissner--Nordstr\"{o}m Hamiltonian expressed in the canonical variables $(b, p_b, c, p_c)$,
\begin{equation}\label{eq:hamiltonian_classic}
    H_{\rm cl} = -\frac{1}{2G\gamma} \left[ \left( b + \frac{\gamma^2}{b} - \frac{\gamma^2 Q^2}{b p_c} \right) p_b + 2 c p_c \right],
\end{equation}
which satisfies $\{b, p_b\} = G\gamma$ and $\{c, p_c\} = 2G\gamma$~\cite{Tibrewala:2012xb, Gambini:2014qta}, they derived an effective quantum--corrected metric that replaces the classical singularity with a smooth transition surface. The quantization is implemented through the polymerization procedure as~\cite{Gambini:2021uzf}
\begin{equation}
b \;\longrightarrow\; \frac{\sin(\delta_b b)}{\delta_b}, \qquad 
p_b \;\longrightarrow\; \frac{p_b}{\cos(\delta_b b)},
\end{equation}
where $\delta_b$ is the (dimensionless) polymerization parameter. Using substitutions in the classical Hamiltonian yields the effective Hamiltonian~\cite{Borges:2023fub}
\begin{align}
\mathcal{H}_{\rm eff} = -\frac{1}{2G\gamma b_0} &\left[ \frac{\sin(\delta_b b)}{\delta_b}
+ \frac{\gamma^2\delta_b}{\sin(\delta_b b)}
- \frac{\gamma^2\delta_b Q^2}{\sin(\delta_b b) \, p_c} \right] p_b \nonumber \\
&\quad - \frac{c \, p_c \cos(\delta_b b)}{b_0},
\end{align}
where we have $b_0 \equiv \sqrt{1 + \gamma^2\delta_b^2}$ and the $Q$ is the electric charge. Solving the resulting Hamilton equations and performing an analytic continuation to the exterior static patch, one obtains the asymptotically flat spacetime metric~\cite{Borges:2023fub}
\begin{equation}\label{eq:lqg_metric}
ds^2 = -A(r)\,dt^2 + \frac{dr^2}{A(r)\left[1 - \dfrac{r_0}{M}\,g(r)\right]} + r^2\,d\Omega^2,
\end{equation}
where
\begin{equation}
A(r) = 1 - \frac{2M}{r} + \frac{Q^2}{r^2}
\end{equation}
is the  Reissner-Nordstr\"{o}m lapse function. The quantum correction factor is encoded through
\begin{equation}
g(r) = \frac{2Mr - Q^2}{r^2} \left(1 + \sqrt{1 - \dfrac{b_0^2 Q^2}{(b_0^2-1)M^2}}\right)^{-1},
\end{equation}
and the transition surface radius $r_0$ is given by
\begin{equation}\label{eq:r0}
r_0 = \frac{(b_0^2 - 1) M}{b_0^2} \left( 1 + \sqrt{1 - \frac{b_0^2 Q^2}{(b_0^2 - 1)^2 M^2}} \right).
\end{equation}
Here, the mass of the black hole $M$ and the electric charge $Q$ remain constants of motion. One can verify that as the polymerization parameter vanishes, i.e., $\delta_b \rightarrow 0$ or equivalently $b_0 \rightarrow 1$, the metric solution~\eqref{eq:lqg_metric} reduces to the standard Reissner--Nordstr\"{o}m black hole. In general, the metric~\eqref{eq:lqg_metric} shares the same horizon radii as the classical Reissner--Nordstr\"{o}m solution,
\begin{equation}
r_{\pm} = M\left(1 \pm \sqrt{1 - Q^2/M^2}\right).
\end{equation}
The interior geometry of the black hole is modified due to the quantum transition surface at the areal radius $r_0$~\eqref{eq:r0}.The classical Reissner--Nordstr\"{o}m singularity at $r = 0$ is absent; instead, the spacetime undergoes a smooth transition to a white hole interior at $r = r_0$.

A crucial consequence of LQG is that the polymerization parameter $\delta_b$ is not a free constant but is fixed by imposing the minimal area condition~\cite{Modesto:2008im, Borges:2023fub},
\begin{equation}
4\pi r_0^2 = 4\pi\sqrt{3}\,\gamma,
\end{equation}
which uniquely determines $\delta_b$ as a function of the mass $M$ and the electric charge $Q$~\cite{Borges:2023fub, Baranov:2024myo}, given by\footnote{Here, we set $\gamma = \sqrt{3}/6$~\cite{Pigozzo:2020zft}, and we adopt units in which $G = c = \ell_{\rm Pl} = 1$.}
\begin{equation}\label{eq:deltab}
\delta_b^2 = \frac{12}{2\sqrt{2}\,M - 2Q^2 - 1}.
\end{equation}
Since $M$ and $Q$ are constants of motion, the polymerization parameter $\delta_b$ is also a constant of motion along a given dynamical trajectory~\cite{Bodendorfer:2019cyv}. From Eq.~\eqref{eq:deltab}, one sees that $\delta_b$ grows as $M$ decreases, reflecting the increasing importance of quantum corrections at smaller scales~\cite{Borges:2023fub}. Combining this constraint with the requirement $|Q| \leq M\sqrt{b_0^2-1}/b_0$, one obtains a bound on the charge~\cite{Borges:2023fub,Kumar:2025ueq}
\begin{equation}
Q^2 \leq \frac{\sqrt{2}}{2}\,M,
\end{equation}
which is more restrictive than the classical extremality condition $|Q| \leq M$. To make the connection between the charge parameter $Q$ and the LQG polymerization parameter explicit, we note that through the minimal area condition~\eqref{eq:deltab}, each value of $Q$ uniquely determines $\delta_b$ and hence $b_0$. Table~\ref{tab:params} lists the parameter values used in this work. For all $Q$ we consider, $b_0$ deviates noticeably from $1$, so every configuration sits in a genuinely quantum--gravity–corrected regime. Even for the neutral case $Q = 0$, we get $b_0 \approx 1.24$, indicating that LQG effects are not merely perturbative and already modify the background geometry itself.

\begin{table}[h]
\centering
\begin{tabular}{cccc}
\hline
$Q$ & $\delta_b^2$ & $b_0$ & $r_0/M$ \\
\hline
$0.0$ & $6.56$ & $1.24$ & $0.71$ \\
$0.2$ & $6.86$ & $1.25$ & $0.69$ \\
$0.4$ & $7.95$ & $1.29$ & $0.65$ \\
$0.6$ & $10.82$ & $1.38$ & $0.66$ \\
$0.8$ & $21.88$ & $1.68$ & $1.08$ \\
\hline
\end{tabular}
\caption{Values of the LQG polymerization parameter $\delta_b$, the 
derived quantity $b_0 = \sqrt{1+\gamma^2\delta_b^2}$, and the transition 
surface radius $r_0$ for each value of the charge parameter $Q$ used 
in this work, computed with $M = 1$ and $\gamma = \sqrt{3}/6$.}
\label{tab:params}
\end{table}

\section{Geodesics and orbits}\label{sec:geodesics_orbits}

Before computing gravitational waveforms, we need to understand the geodesic motion of a test particle in the metric~\eqref{eq:lqg_metric}. We first derive the effective potential and locate the ISCO and the MBO. Then we classify periodic orbits using the Levin--Perez--Giz taxonomy \cite{Levin:2008mq}.

\subsection{Timelike geodesics}
\label{subsec:geodesics}

For a massive test particle in the charged LQG black hole~\eqref{eq:lqg_metric}, we derive the geodesic equations using the Lagrangian~\cite{Chandrasekhar:1985kt, Wald:1984rg, Misner:1973prb} approach. We restrict our analysis to the equatorial plane $\theta = \pi/2$ ($\dot{\theta}=0$) and rescaling coordinates as $r\to r/M$, $r_0\to r_0/M$, $Q\to Q/M$, $t\to t/M$, we have
\begin{equation}
2\mathcal{L} = -A(r)\,\dot{t}^2 + \frac{\dot{r}^2}{A(r)\!\left[1 - r_0 g(r)\right]} + r^2\dot{\phi}^2,
\label{eq:lagrangian}
\end{equation}
where dots denote derivatives with respect to the affine parameter $\tau$.

Plugging the Lagrangian into the Euler--Lagrange equations gives two conserved quantities,
\begin{equation}
E = A(r)\,\dot{t}, \qquad L = r^2\dot{\phi},
\label{eq:conserved}
\end{equation}
which come from the Killing vectors $\partial_t$ and $\partial_\phi$, and they reflect time--translation and rotational symmetry~\cite{Chandrasekhar:1985kt, Wald:1984rg}. Here $E$ and $L$ are the specific energy and angular momentum of the test particle. If we substitute these back into the Lagrangian and enforce $2\mathcal{L} = -1$ for timelike motion, we get the radial equation
\begin{equation}\label{eq:radial}
\dot{r}^2 = \bigl(E^2 - V_{\rm eff}(r)\bigr)\bigl[1 - r_0 g(r)\bigr],
\end{equation}
where
\begin{equation}
V_{\rm eff}(r) = A(r)\left(1 + \frac{L^2}{r^2}\right) = \left(1 - \frac{2}{r} + \frac{Q^2}{r^2}\right)\!\left(1 + \frac{L^2}{r^2}\right).
\label{eq:Veff}
\end{equation}
Far away from the black hole, $r \to \infty$, we get $V_{\rm eff} \to 1$. That's what we expect for a flat spacetime, and it means $E$ is just the energy per unit mass measured by an observer at infinity~\cite{Chandrasekhar:1985kt, Wald:1984rg}.

The value $E = 1$ marks the boundary between bound and unbound motion. If $E > 1$, the particle is unbound and can fly off to infinity. If $E < 1$, it follows a bound trajectory. Stable bound orbits sit between the ISCO and the MBO, so a particle on a bound orbit has $E_{\rm ISCO} \leq E \leq E_{\rm MBO} = 1$ and $L \geq L_{\rm ISCO}$~\cite{Nasereldin:2019afz, Chandrasekhar:1985kt}.

Working with $\dot{r}^2$ directly can be annoying because of the sign ambiguity when the particle turns around. Better to use the second--order radial equation. Differentiating~\eqref{eq:radial} with respect to $\tau$ gives
\begin{equation}\label{eq:radial2}
\ddot{r} = -\frac{1}{2}\frac{dV_{\rm eff}}{dr}\!\left[1 - g(r)\right]
- \frac{1}{2}\!\left(E^2 - V_{\rm eff}\right) r_0\frac{dg(r)}{dr}.
\end{equation}
This is much more convenient for numerical computations, since there is no need to track sign changes at the turning points, which allows a smooth description of the radial motion throughout the orbit~\cite{Levin:2008mq, Glampedakis:2002cb, Misner:1973prb}.

\subsection{Marginally bound and innermost stable circular orbits}
\label{subsec:mbo_isco}

Circular orbits are defined by $\dot{r} = 0$ and $\ddot{r} = 0$, which from~\eqref{eq:radial} and~\eqref{eq:radial2} give
\begin{equation}\label{eq:circular_cond}
V_{\rm eff}(r) = E^2, \qquad \frac{dV_{\rm eff}(r)}{dr} = 0.
\end{equation}

\begin{figure}
    \centering
    \subfigure{\includegraphics[width=0.9\linewidth]{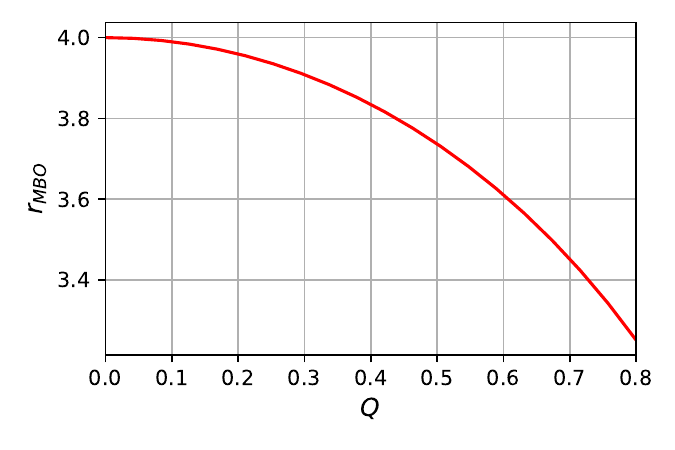}}
    \subfigure{\includegraphics[width=0.9\linewidth]{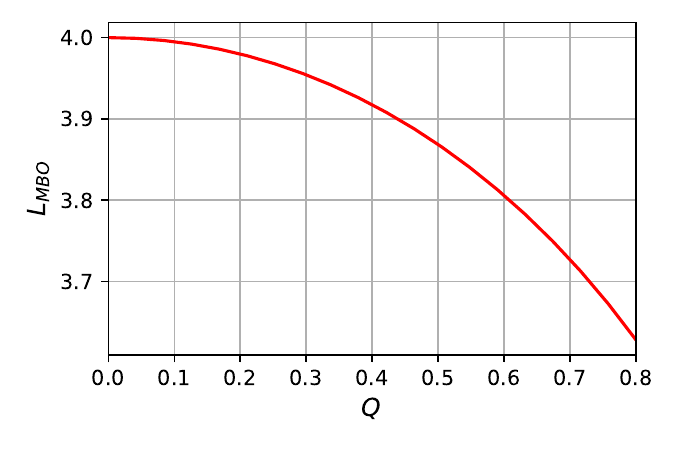}}
    \caption{The figure displays the behavior of $r_{MBO}$ (upper panel) and $L_{MBO}$ (lower panel) as a function of the charge $Q$.}
    \label{fig:mbo}
\end{figure}

The MBO is the innermost circular orbit where the particle remains marginally bound, so $E_{\rm MBO} = 1$. Its radius $r_{\rm MBO}$ and angular momentum $L_{\rm MBO}$ come from solving~\cite{Chandrasekhar:1985kt, Bardeen:1972fi}
\begin{align}
V_{\rm eff}(r)\big|_{r_{\rm MBO}} &= 1, \\
\frac{dV_{\rm eff}(r)}{dr}\bigg|_{r_{\rm MBO}} &= 0.
\end{align}
Since the metric has LQG corrections, we have to solve them numerically. Figure~\ref{fig:mbo} shows $r_{\rm MBO}$ and $L_{\rm MBO}$ versus $Q$, both decrease as $Q$ increases. The quantum corrections effectively pull the MBO inward and reduce the angular momentum needed for a bound orbit compared to the classical Reissner--Nordström case.

\begin{figure}
    \centering
    \subfigure{\includegraphics[width=0.9\linewidth]{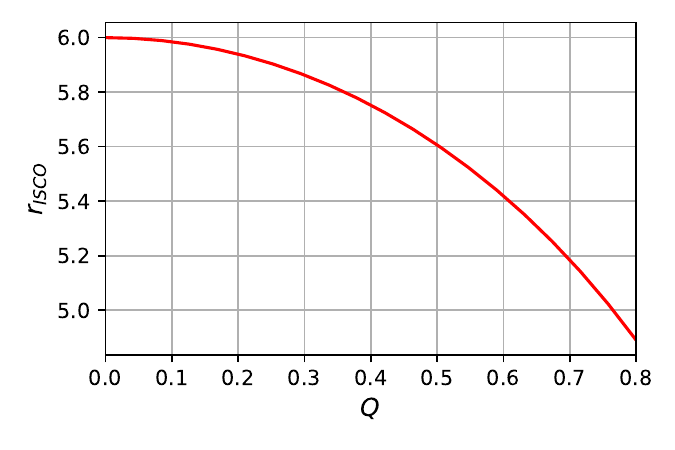}}
    \subfigure{\includegraphics[width=0.9\linewidth]{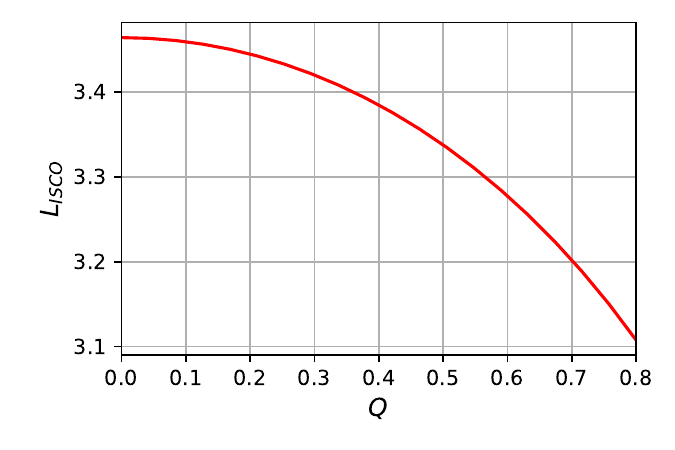}}
    \subfigure{\includegraphics[width=0.9\linewidth]{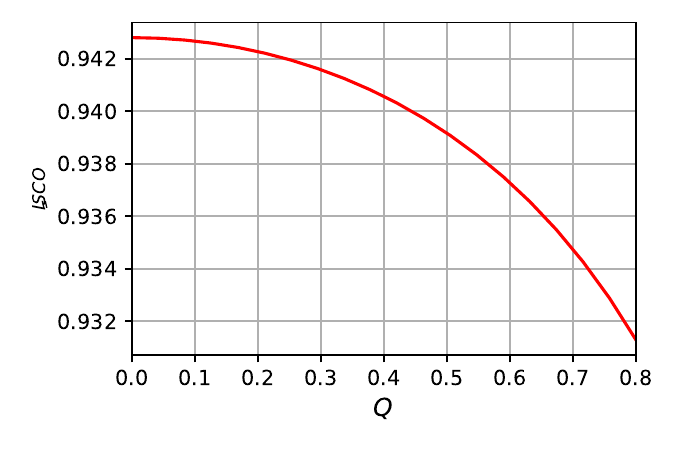}}
    \caption{The figure displays the behavior of $r_{ISCO}$ (upper panel), $L_{ISCO}$ (middle panel), and $E_{ISCO}$ (the bottom panel) as a function of the charge $Q$.}
    \label{fig:isco}
\end{figure}

The ISCO marks the boundary between stable and unstable circular motion. It's a key quantity in black hole accretion physics~\cite{Bardeen:1972fi, Chandrasekhar:1985kt, Wald:1984rg}. Inside $r_{\rm ISCO}$, circular orbits are unstable, and any infalling material just plunges into the black hole. The ISCO parameters come from three conditions:
\begin{align}
V_{\rm eff}(r)\big|_{r_{\rm ISCO}} &= E_{\rm ISCO}^2, \label{eq:isco1}\\
\frac{dV_{\rm eff}(r)}{dr}\bigg|_{r_{\rm ISCO}} &= 0, \label{eq:isco2}\\
\frac{d^2 V_{\rm eff}(r)}{dr^2}\bigg|_{r_{\rm ISCO}} &= 0, \label{eq:isco3}
\end{align}
The first two enforce circular motion and the third picks out the marginally stable one~\cite{Chandrasekhar:1985kt}. Figure~\ref{fig:isco} shows our numerical results for $r_{\rm ISCO}$, $E_{\rm ISCO}$, and $L_{\rm ISCO}$ versus $Q$. All three decrease when we increase the charge.

\begin{figure}
    \centering
    \includegraphics[width=0.9\linewidth]{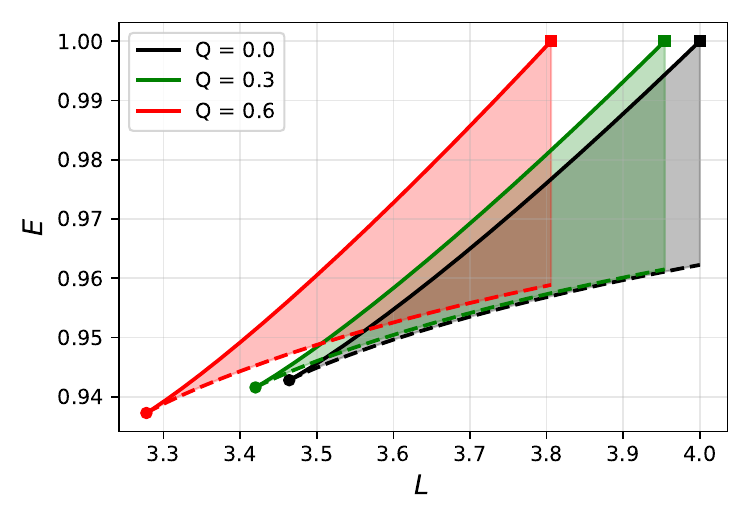}
    \caption{The shaded colored regions show the allowed parameter space $(E, L)$ for the time--like orbit around the charged LQG black hole for different values of the charge $Q$.}
    \label{fig:EL}
\end{figure}

Stable bound circular orbits only exist for angular momenta between $L_{\rm ISCO}$ and $L_{\rm MBO}$~\cite{Chandrasekhar:1985kt, Bardeen:1972fi, Wald:1984rg}. So it's handy to parametrize
\begin{equation}\label{eq:L_param}
L = L_{\rm ISCO} + \epsilon\,(L_{\rm MBO} - L_{\rm ISCO}),
\end{equation}
with $0 \leq \epsilon \leq 1$. Here $\epsilon = 0$ is the ISCO and $\epsilon = 1$ is the MBO. This keeps us in the physically relevant region of bound orbits.
Figure~\ref{fig:EL} shows the allowed $(E, L)$ parameter space for bound timelike orbits at different $Q$. The specific energy $E$ increases with $L$ increases, as expected expect for circular geodesics in spherical spacetimes~\cite{Chandrasekhar:1985kt}. Also, as $Q$ gets larger, the allowed $(E, L)$ region widens.

\subsection{Periodic Orbits}
\label{subsec:periodic}

Periodic orbits constitute a discrete, yet dense, subset of all bound timelike orbits and are of particular astrophysical relevance because they produce quasi--periodic gravitational wave signals observable by space--based detectors~\cite{Levin:2008mq, Glampedakis:2002cb, Barack:2003fp}. In general, the radial and azimuthal motions of a bound orbit have incommensurable frequencies, producing non--closing, ergodic trajectories. Periodicity is achieved when the ratio of these frequencies is rational.
Levin and Perez--Giz~\cite{Levin:2008mq} introduced a systematic taxonomy wherein we can label each periodic orbit uniquely by a triplet of non--negative integers $(z, w, v)$, representing the number of zooms, whirls, and vertices completed during one full radial period. The essential dimensionless parameter encoding the orbital geometry is the frequency ratio
\begin{equation}
q = \frac{\omega_\phi}{\omega_r} - 1 \equiv w + \frac{v}{z},
\label{eq:q}
\end{equation}
where $\omega_\phi$ and $\omega_r$ are the azimuthal and radial angular frequencies, respectively. The quantity $q$ measures the periapsis advance accumulated per radial oscillation beyond that of a simple closed ellipse, and it encodes the geometric structure of the orbit in the strong--field regime~\cite{Levin:2008mq, Glampedakis:2002cb}.

For a periodic orbit, the total azimuthal angle swept out during one full radial period has to be an integer multiple of $2\pi$:
\begin{equation}\label{eq:deltaphi}
\frac{\omega_\phi}{\omega_r} = \frac{\Delta\phi}{2\pi}, \qquad \Delta\phi = 2\int_{r_1}^{r_2}\frac{\dot{\phi}}{\dot{r}}\,dr,
\end{equation}
where $r_1$ and $r_2$ are the periapsis and apoapsis radii (closest and farthest points). In other words, the orbit closes after a certain number of radial oscillations only if the ratio of the angular frequencies is rational.
Plugging in the equations of motion~\eqref{eq:conserved} and~\eqref{eq:radial} gives an expression for the frequency ratio
\begin{equation}\label{eq:q_integral}
q = \frac{L}{\pi}\int_{r_1}^{r_2}
\frac{dr}{r^2\sqrt{\left(1 - \dfrac{r_0}{M}g(r) \right) \; \big(E^2 - V_{\rm eff}(r) \big)}}\, - 1.
\end{equation}
The quantity $q$ is what Levin and Perez--Giz call the \enquote{zoom--whirl} parameter. It tells you how much the orbit precesses per radial cycle beyond a simple ellipse. If $q$ is rational, the orbit is periodic; if it's irrational, the orbit never quite closes and eventually fills up a region of space.
We evaluate this integral numerically for the LQG metric~\eqref{eq:lqg_metric}. For a fixed set of parameters $(M, Q, \delta_b)$, we find an orbit with topological label $(z, w, v)$ by searching for the $E$ and $L$ values (within the bound--orbit region from~\eqref{eq:L_param}) that make $q$ equal to the rational number $w + v/z$~\cite{Levin:2008mq, Glampedakis:2002cb, Grossman:2008yk}. Once such an orbit is identified, the numbers $(z, w, v)$ completely characterize its structure, including the number of zooms, whirls, and vertices.

\begin{figure}
    \centering
    \subfigure{\includegraphics[width=0.93\linewidth]{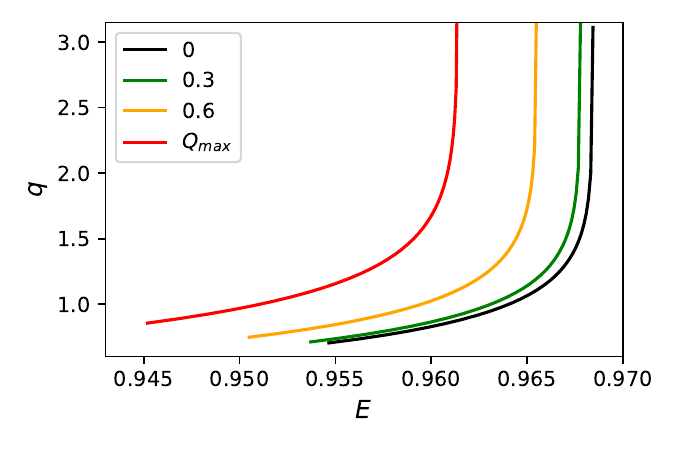}}
    \subfigure{\includegraphics[width=0.9\linewidth]{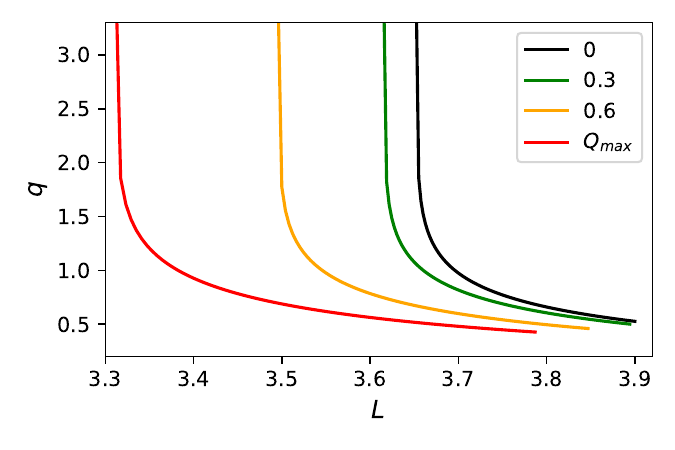}}
    \caption{The plot presents the behavior of the rational frequency ratio $q = \omega_\phi / \omega_r - 1$ as a function of energy $E$ (upper panel), and as a function of angular momentum $L$ for fixed energy $E = 0.96$ (lower panel). the dimension less parameter $\epsilon$ is taken as $\epsilon = 0.5$, and the plot shows the frequency ratio for different value of the charge parameter $Q$.}
    \label{fig:qLE}
\end{figure}
\begin{figure*}
    \centering
    \includegraphics[width=1\linewidth]{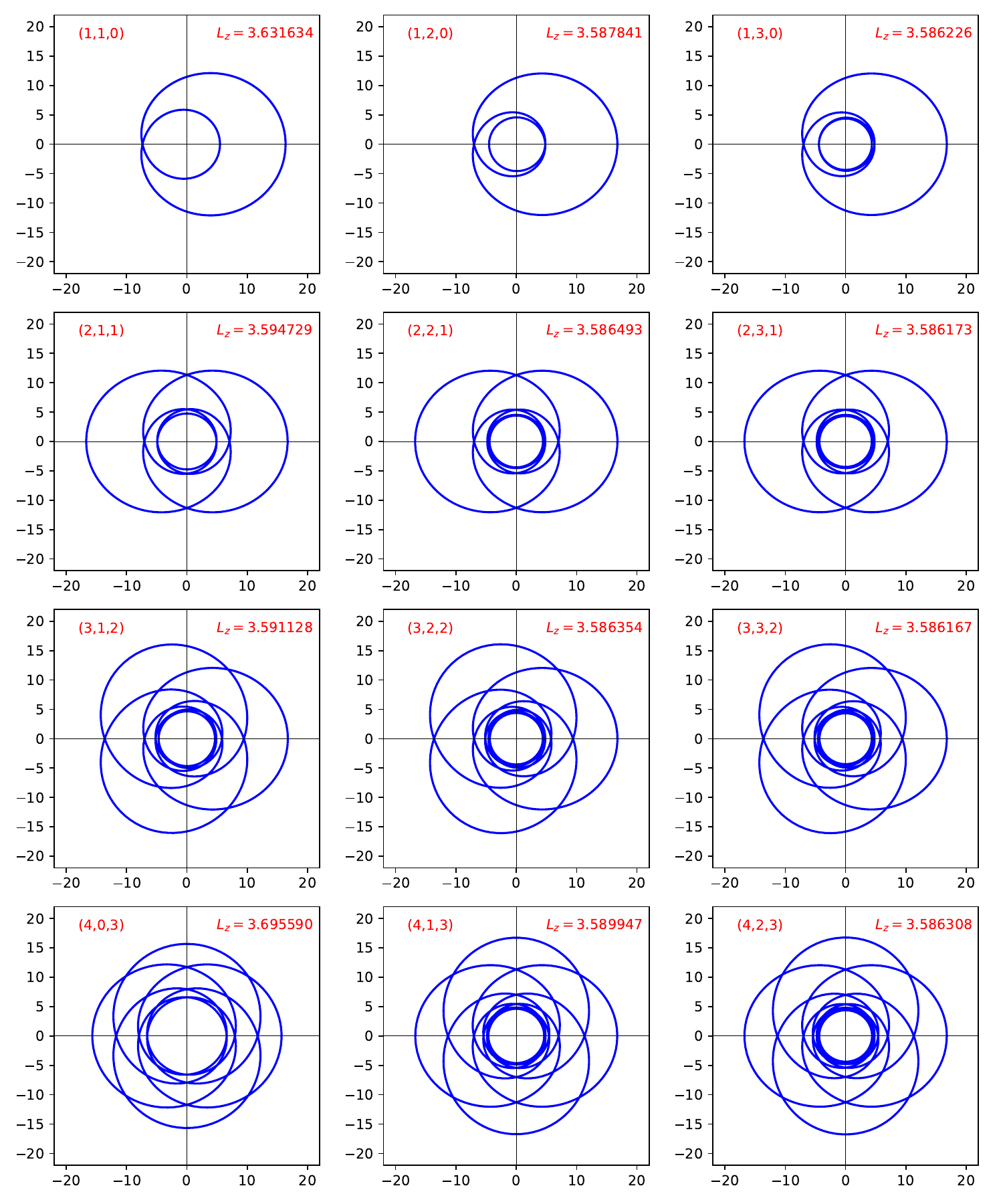}
    \caption{The plot displays different periodic orbits around the LQG black hole classified by the topological non-negative integer $(z, w, v)$ for energy $E = 0.96$, and the charge $Q = 0.4$. Each panel shows the priodic orbits for different values of zoom ($z$), whirl ($w$), and vertex ($v$) numbers in which the corresponding angular momentum is also indicated in the plot. More complex orbital structure are generated as the zoom number increases, while more loop around the black hole is predicted by increasing the whirl number.}
    \label{fig:orbits}
\end{figure*}

Figure~\ref{fig:qLE} shows how $q$ depends on the specific energy $E$ (upper panel, with $\epsilon = 0.5$ fixed) and on the angular momentum $L$ (lower panel, at fixed $E$), for a few different values of the charge $Q$.
Take the upper panel first, as $E$ increases toward its maximum allowed value for a bound orbit, $q$ grows steadily.
This behavior is expected, since relativistic periapsis precession becomes stronger as the orbit approaches the separatrix.~\cite{Levin:2008mq, Glampedakis:2002cb}. Notice also that the maximum possible $E$ gets smaller when $Q$ is larger. So a higher charge actually lowers the energy ceiling for bound orbits.
Now look at the lower panel, here, $q$ drops as $L$ increases. Larger angular momentum pushes the orbit outward, so precession weakens. The frequency ratio $q$ keeps decreasing until $L$ hits its minimum allowed value at the ISCO~\cite{Chandrasekhar:1985kt, Bardeen:1972fi}.
Figure~\ref{fig:orbits} shows sample periodic orbits for several $(z, w, v)$ triplets. We get these by numerically integrating~\eqref{eq:radial2} together with $\dot{\phi} = L/r^2$, starting from apoapsis with $\dot{r}=0$. As 
one expects from the Levin--Perez--Giz scheme~\cite{Levin:2008mq, Grossman:2008yk}, bigger $z$ gives more complex multi--leaf patterns, while bigger $w$ adds extra near--circular loops around the black hole before each radial return. This growing complexity is a telltale sign of strong--field relativistic dynamics near the ISCO~\cite{Bambi:2017khi, Haroon:2025rzx}.

\begin{table*}[htb!]
 {\footnotesize
\resizebox{\textwidth}{!}{ 
 \begin{centering}	
	\begin{tabular}{|c c c c c c c c c c c c c c|}

\hline
{$Q$ } & {$L$}& {$E_{(1,1,0)}$}& {$E_{(1,2,0)}$} &{$E_{(1,3,0)}$} & {$E_{(2,1,1)}$} & {$E_{(2,2,1)}$} &{$E_{(2,3,1)}$} & {$E_{(3,1,2)}$ }& {$E_{(3,2,2)}$} &{$E_{(3,3,2)}$} &{$E_{(4,0,3)}$ } &{$E_{(4,1,3)}$ }&{$E_{(4,2,3)}$} \\ \hline
$0$&$3.7320$& $0.9654$ &$0.9683$&$0.9684$& $0.9680$& $0.9684$&$0.9684$ & $0.9682$& $0.9684$&$0.9684$ &$0.9599$ & $0.9682$&$0.9684$\\
\hline
$0.2$&$3.7122$& $0.9636$ &$0.9680$& $0.9681$& $0.9674$& $0.9681$& $0.9682$& $0.9677$ &$0.9681$&$0.9682$& $0.9564$& $0.9678$&$0.9681$ \\

\hline
$0.4$&$3.6510$& $0.9623$ &$0.9671$& $0.9672$& $0.9664$& $0.9672$& $0.9673$ & $0.9667$& $0.9672$ &$0.9673$& $0.9545$ & $0.9669$&$0.9672$\\
\hline
{$0.6$}&$3.5417$& $0.9594$ &$0.9652$& $0.9654$& $0.9644$& $0.9654$&$0.9655$& $0.9648$ &$0.9655$&$0.9655$& $0.9506$& $0.9650$&$0.9654$ \\
\hline
{$0.8$}&$3.3679$& $0.9533$ &$0.9619$& $0.9623$& $0.9604$& $0.9622$& $0.9623$& $0.9611$ &$0.9623$&$0.9623$& $0.9458$& $0.9614$&$0.9623$ \\
       \hline
	\end{tabular} 
\end{centering}}
	\caption{Numerical results of the energy for different periodic orbits around LQG black holes for different values of the charge $Q$. We kept $\epsilon=0.5$ for these results.}\label{tab:energies}
    }
\end{table*}

Table~\ref{tab:energies} lists the specific energies corresponding to the first several orbit families $(z, w, v)$ for representative values of $Q$, with $\epsilon = 0.5$ fixed. The tabulated values demonstrate that the LQG polymerization parameter and hence, the black hole charge produce systematic shifts in orbital energies, which could in principle be imprinted on the gravitational wave spectrum emitted by an extreme--mass--ratio inspiral system.

\section{Gravitational Waves}\label{sec:gw}

Among the most promising targets for next--generation space--based gravitational wave detectors, LISA~\cite{Danzmann:1997hm, Schutz:1999xj, Gair:2004iv, LISA:2017pwj, Maselli:2021men}, Taiji~\cite{Hu:2017mde}, and TianQin~\cite{TianQin:2015yph, Gong:2021gvw}, are EMRIs. These systems pair a stellar--mass compact object, a black hole or neutron star of mass roughly $1$--$100\,M_\odot$, with a supermassive black hole of mass $\sim 10^6$--$10^7\,M_\odot$~\cite{LISA:2017pwj}. The small body orbits the big one on a long--lived path. Because the mass ratio is so extreme, only a tiny fraction of the orbital energy gets radiated away each cycle. So the inspiral is slow, and the system hangs around in the millihertz band for many cycles; right where LISA, Taiji, and TianQin are most sensitive~\cite{Babak:2017tow, Barack:2018yvs}.
EMRIs are not just observationally promising; they're also theoretically powerful. The gravitational waves they emit carry detailed information about the strong--field dynamics and the spacetime geometry of the central object~\cite{Barack:2003fp, Gair:2012nm, Glampedakis:2002cb}. In recent years, EMRIs have been proposed as probes of modified gravity, dark matter environments, and even quantum--gravity effects~\cite{Chen:2026kbn, Fu:2024cfk, Yang:2025esa}. In our case, the central object is the charged LQG black hole from metric~\eqref{eq:lqg_metric}. Both the polymerization parameter $\delta_b$ and the charge $Q$ alter the geodesic structure and orbital frequencies compared to the classical Reissner--Nordström case. These changes show up in the emitted gravitational waves. That means future high--precision measurements could either detect these LQG signatures or at least put meaningful bounds on the parameters~\cite{Torres-Orjuela:2024gux}.

\subsection{Adiabatic Approximation and the Numerical Kludge Method}

Computing waveforms for a charged LQG black hole EMRI is not trivial and needs a sensible approximation strategy. We work in the adiabatic approximation, which holds when the mass ratio is tiny, $m/M \ll 1$~\cite{Mino:1997bx, Poisson:2011nh, Barack:2009ux}. In this regime, radiation reaction acts much more slowly than the orbital period. So over a single orbit, the small body's energy and angular momentum barely change. That means we can describe the motion as a sequence of geodesics in a fixed background spacetime~\cite{Strateny:2025zkd}. We also neglect self--force backreaction entirely. The companion is treated as a point--like test particle following exact geodesics of the LQG metric. 
Although self-force effects are important for constructing high-precision matched-filter templates, neglecting them is adequate for capturing the qualitative features of the orbital dynamics and waveform morphology, and is standard practice in exploratory studies of strong-field systems~\cite{Barack:2009ux, Poisson:2011nh, Cardenas-Avendano:2024mqp}.

To actually build the waveforms, we use the numerical kludge (NK) method from Babak et al.~\cite{Babak:2006uv, Gair:2005ih}. This is a semi--relativistic approach: we integrate the exact geodesics in the curved background, then plug the motion into the flat--spacetime quadrupole formula. It's not as rigorous as a full Teukolsky treatment, but it captures the key strong--field features of EMRI waveforms, including the zoom--whirl structure, at a fraction of the computational cost. The NK method has been used widely to study orbital dynamics and beyond--GR or environmental effects in gravitational wave signals~\cite{Babak:2006uv, Gair:2005ih, Qiao:2024gfb, Alloqulov:2025bxh, Haroon:2025rzx, Ahmed:2025azu}.  That quadrupole--based fluxes remain in agreement with full Teukolsky results across a broad range of orbital configurations~\cite{Strateny:2025zkd}, and that LISA--class detectors can potentially distinguish quantum gravity modifications with parameter constraints as tight as $10^{-6}$~\cite{Zi:2024jla, Chen:2026kbn}.

\subsection{Waveform Construction}
The NK procedure starts with the geodesic equations, specifically,~\eqref{eq:radial2} and $\dot{\phi} = L/r^2$, which we integrate numerically to get the particle's trajectory $Z^i(t)$ in the background spacetime. We solve these equations in spherical coordinates $(r, \theta, \phi)$. Since the motion stays in the equatorial plane ($\theta = \pi/2$), we then map the trajectory to a Cartesian frame centered on the black hole~\cite{Babak:2006uv}:
\begin{equation}
    x = r\sin\theta\cos\phi, \quad y = r\sin\theta\sin\phi, \quad z = r\cos\theta. 
\end{equation}
The leads to the particle's position $x_i(t)$, velocity $v_i(t) = \dot{x}_i(t)$, and acceleration $a_i(t) = \ddot{x}_i(t)$ in Cartesian coordinates. 

We work in natural units $G = c = 1$. The gravitational wave metric perturbation is related to the symmetric trace--free (STF) mass quadrupole moment $I_{ij}^{\rm STF}$ by the standard formula~\cite{Maggiore:2007ulw, Misner:1973prb, Thorne:1980ru}
\begin{equation}\label{hij_Iij}
    h_{ij} = \frac{2}{D_L}\,\ddot{I}_{ij}^{\rm STF},
\end{equation}
where $D_L$ is the luminosity distance to the source. This is the quadrupole approximation: the leading--order contribution to gravitational radiation comes from the second time derivative of the mass quadrupole. Higher multipoles (octupole, etc.) are suppressed by factors of $v/c$ and matter less for our purposes.
For a point particle of mass $m$ at position $Z^i(t)$, the mass quadrupole moment is
\begin{equation}\label{Iij}
    I_{ij} = m\int d^3x\,x^i x^j\,\delta^3\!\big(x^i - Z^i(t)\big) = m\,x_i x_j.
\end{equation}
In other words, it's just the particle's mass times the outer product of its position vector with itself.

Plugging $I_{ij} = m\,x_i x_j$ into~\eqref{hij_Iij} and taking the two time derivatives gives~\cite{Maggiore:2007ulw, Babak:2006uv, Thorne:1980ru}
\begin{equation}
    h_{ij} = \frac{2m}{D_L}\!\left(a_i x_j + a_j x_i + 2v_i v_j\right)^{\rm STF},
\end{equation}
where $v_i = \dot{x}_i$ and $a_i = \ddot{x}_i$ are the Cartesian velocity and acceleration components we get directly from our numerical integration. The STF projection ensures that we only keep the traceless part of the expression, which is the part that actually radiates. This is a standard result: the gravitational wave strain is proportional to the second time derivative of the quadrupole moment, and for a point mass in motion, that derivative works out to the combination of positions, velocities, and accelerations shown above.

\begin{figure}[!ht]
    \centering
    \subfigure{\includegraphics[width=0.9\linewidth]{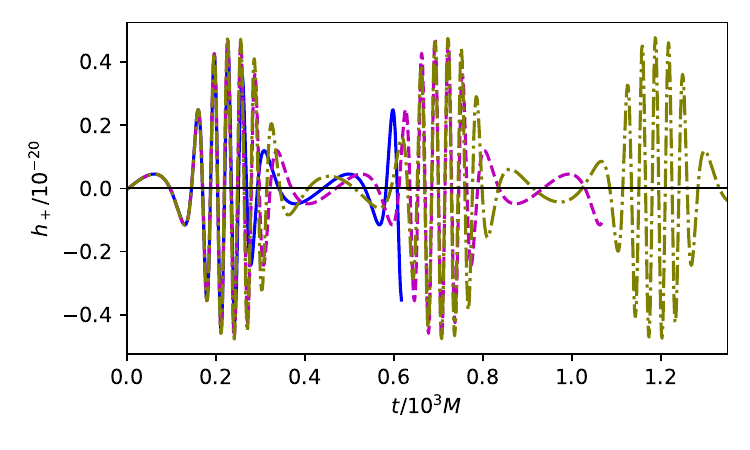}}
    \subfigure{\includegraphics[width=0.9\linewidth]{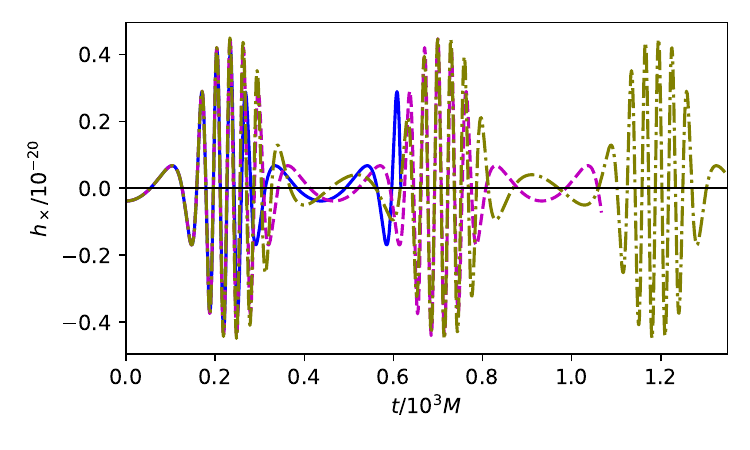}}
    \caption{GW polarizations $h_+$ (top) and $h_\times$ (bottom) for three periodic orbits: $(z,w,v) = (1,2,0)$ (blue), $(2,2,1)$ (magenta), and $(3,2,2)$ (olive). All curves are for $Q = 0.4$ and $E = 0.96$. The zoom phase produces smooth, low--amplitude oscillations, while the whirl phase gives sharp, high-amplitude bursts. 
    The system is taken as a smaller-mass object, $m = 10 M_\odot$, moving around a supermassive LQG black hole with mass $M = 10^6 M_\odot$, where inclination $\iota = \zeta = \pi/4$ and luminosity distance is taken as $D_L = 200 \; {\rm Mpc}$}
    \label{fig:hchp_zwv}
\end{figure}

\subsection{Extraction of the Gravitational Wave Polarisations}
To relate the source--frame metric perturbation $h_{ij}$ to the observable polarizations, we need to project onto the detector frame. The transformation from the source frame to the detector frame uses two angles~\cite{Babak:2006uv, Maggiore:2007ulw}:
\begin{itemize}
    \item $\iota$ (inclination) — the angle between the orbital angular momentum and the line of sight to the observer.
    \item $\zeta$ (longitude of pericenter) — the orientation of the orbit's pericenter relative to the detector.
\end{itemize}
In terms of these angles, the orthonormal basis vectors of the detector frame (expressed in source Cartesian coordinates) are
\begin{align}
\hat{e}_X &= (\cos\zeta,\,-\sin\zeta,\,0),\\
\hat{e}_Y &= (\sin\iota\sin\zeta,\,\cos\iota\cos\zeta,\,-\sin\iota),\\
\hat{e}_Z &= (\sin\iota\sin\zeta,\,-\sin\iota\cos\zeta,\,\cos\iota).
\end{align}
Projecting $h_{ij}$ onto these basis vectors gives the two physical GW polarizations~\cite{Babak:2006uv, Gair:2007uk}:
\begin{align}\label{h_plus_h_cross}
h_+ &= \frac{1}{2}\!\left(e_X^i e_X^j - e_Y^i e_Y^j\right)h_{ij},\\
h_{\times} &= \frac{1}{2}\!\left(e_X^i e_Y^j + e_Y^i e_X^j\right)h_{ij}.
\end{align}
The plus polarization $h_+$ describes the wave's stretching and squeezing along the $X$ and $Y$ axes, while the cross polarization $h_{\times}$ does the same but rotated by $45^\circ$~\cite{Cardenas-Avendano:2024mqp}.

Following Babak et al.~\cite{Babak:2006uv} and the extension to generic orbits in~\cite{Canizares:2012is}, we define
\begin{align}
h_{\zeta\zeta} &= h_{xx}\cos^2\!\zeta - h_{xy}\sin 2\zeta + h_{yy}\sin^2\!\zeta,\\
\begin{split}
h_{\iota\iota} &= \cos^2\!\iota\!\left[h_{xx}\sin^2\!\zeta + h_{xy}\sin 2\zeta + h_{yy}\cos^2\!\zeta\right] \\
&\quad + h_{zz}\sin^2\!\iota - \sin 2\iota\!\left[h_{xz}\sin\zeta + h_{yz}\cos\zeta\right],
\end{split}\\
\begin{split}
h_{\iota\zeta} &= \frac{1}{2}\cos\iota\!\left[h_{xx}\sin 2\zeta + 2h_{xy}\cos 2\zeta - h_{yy}\sin 2\zeta\right] \\
&\quad + \sin\iota\!\left[h_{yz}\sin\zeta - h_{xz}\cos\zeta\right].
\end{split}
\end{align}
These combinations are just projections of $h_{ij}$ onto directions aligned with $\zeta$ and $\iota$. In terms of them, the polarizations become nice and compact~\cite{Babak:2006uv, Oliver:2023xan}:
\begin{align}\label{hplus_hcross}
h_+ &= \frac{1}{2}\!\left(h_{\zeta\zeta} - h_{\iota\iota}\right),\\
h_{\times} &= h_{\iota\zeta}.
\end{align}
So once we've computed $h_{ij}$ from the particle's motion, we just plug into these formulas and out come the gravitational wave waveforms. For space--based detectors like LISA, Taiji, and TianQin, these polarizations form the basis for matched--filter searches and parameter estimation studies~\cite{Niu:2019ywx, Muguruza:2026hqn}.
\begin{figure*}[!ht]
    \centering
    \begin{tabular}{ c }
        \includegraphics[width=0.4\textwidth]{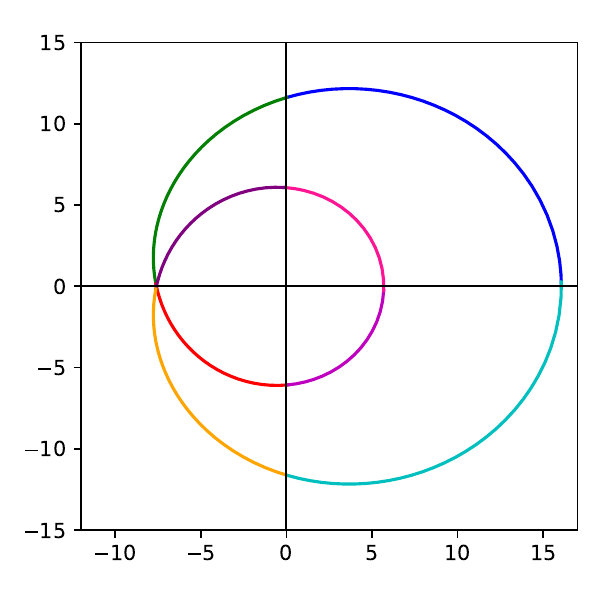}
    \end{tabular}%
    \begin{tabular}{ c c }
        \includegraphics[width=0.45\textwidth]{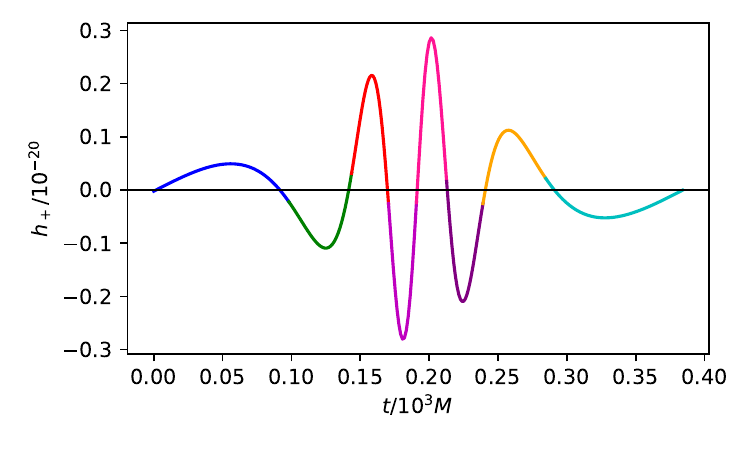}\\ 
        \includegraphics[width=0.45\textwidth]{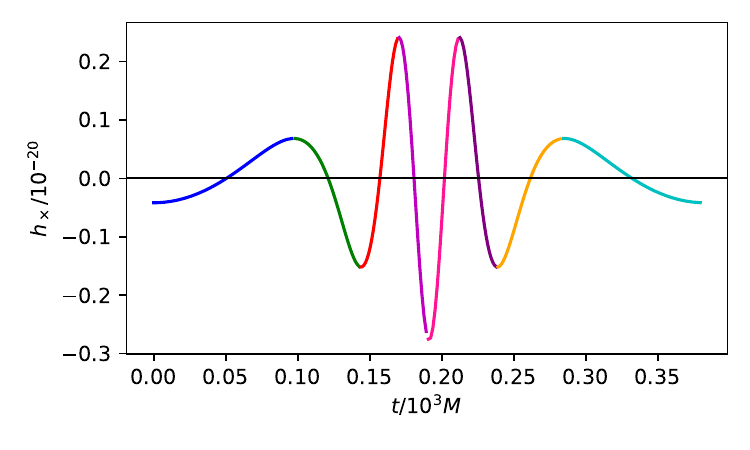} 
    \end{tabular}
    \caption{The correlation between the orbital segments and the waveform is shown for $(1, 1, 0)$ periodic orbit with energy $E = 0.96$ and the charge $Q = 0.4$. different color in the left panel shows the position of the trajectory, and the corresponding polarization waveforms is displayed with the same color on the right panel. The figure shows a map from the specific orbital phase to features in the GW signal. }
    \label{fig:gw_sector}
\end{figure*}

\subsection{Waveform Results: Zoom--Whirl Structure}

We now present our numerical results for the GW polarizations. Throughout this section, we fix the EMRI parameters as follows: the small body has mass $m = 10\,M_\odot$, the central black hole has mass $M = 10^6\,M_\odot$, the inclination and pericenter longitude are $\iota = \zeta = \pi/4$, and the luminosity distance is $D_L = 200\,\mathrm{Mpc}$. These values are typical for LISA--band EMRIs~\cite{LISA:2022yao, Babak:2017tow, Yang:2024lmj}.

Figure~\ref{fig:hchp_zwv} shows $h_+$ and $h_\times$ for three representative periodic orbits: $(z,w,v) = (1,2,0)$, $(2,2,1)$, and $(3,2,2)$. We fix $Q = 0.4$ and $E = 0.96$ for all three. 
Each waveform has the characteristic zoom--whirl structure, a well--known signature of strong--field EMRIs~\cite{Levin:2008mq, Barack:2003fp, Babak:2006uv, Gair:2004iv}. This structure has recently been studied in various quantum--corrected black hole spacetimes~\cite{Ahmed:2025shr, Chen:2026kbn, Yang:2024cnd}. The pattern follows the two phases of the orbit over one radial cycle. During the zoom phase, the particle moves along the large--radius, high--eccentricity part of its orbit, which shows up in the waveform as smooth, low--amplitude oscillations. Then, when the particle swings through periapsis, it enters the whirl phase and executes several rapid, nearly circular loops in the strong--field region near the black hole. The high orbital speed and strong curvature there produce short, high--amplitude bursts in both polarizations~\cite{Glampedakis:2002cb, Grossman:2008yk, Yang:2024lmj}.

So each full radial cycle gives us long stretches of quiet oscillations punctuated by sharp, loud bursts. The number of bursts per cycle equals the whirl number $w$, exactly as the Levin-Perez-Giz taxonomy predicts~\cite{Levin:2008mq, Haroon:2025rzx}. 
As shown in the figure, the $(1,2,0)$, $(2,2,1)$, and $(3,2,2)$ orbits each produce two bursts, consistent with their common whirl number $w=2$. The differences are in how the bursts are spaced and shaped, which depend on $z$ and $v$. Recent work has shown that such periodic orbit waveforms can distinguish between different black hole models~\cite{Chen:2026kbn, Yang:2024lmj}, and that even small quantum corrections produce measurable effects in EMRI waveforms~\cite{Yang:2024cnd, Xamidov:2025oqx}.

To make this orbit--waveform correspondence explicit, Fig.~\ref{fig:gw_sector} shows the $(1,1,0)$ orbit alongside its $h_+$ and $h_\times$ waveforms, with matching colors. This way, each part of the trajectory can be directly linked to the corresponding feature in the waveform — you can see exactly which segment of the orbit produces which wiggle in the signal.


Now let's focus on how the charge parameter $Q$ affects the waveforms. We fix the orbit at $(z,w,v) = (3,2,1)$ and compare results for different $Q$ values, including the neutral case $Q = 0$. The orbits and polarizations are shown in Fig.~\ref{fig:orbitGW_Q}.

The LQG corrections produce two clearly identifiable modifications to the waveform:
\begin{figure*}
    \centering
    \begin{tabular}{ c }
        \includegraphics[width=0.4\textwidth]{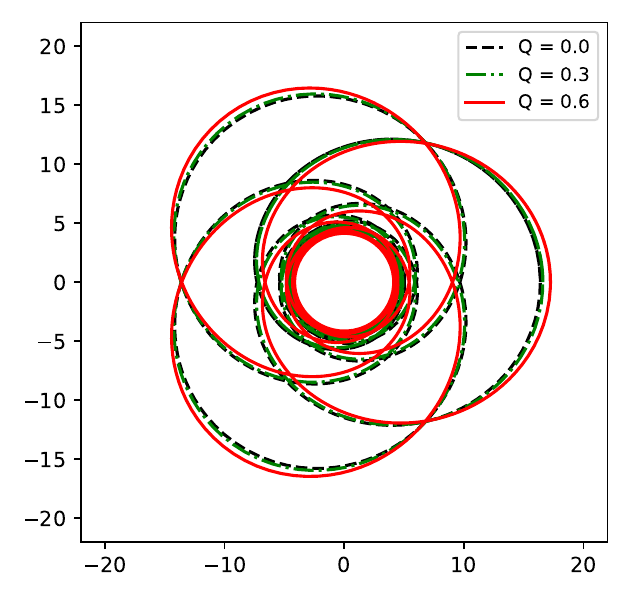}
    \end{tabular}%
    \begin{tabular}{ c c }
        \includegraphics[width=0.45\textwidth]{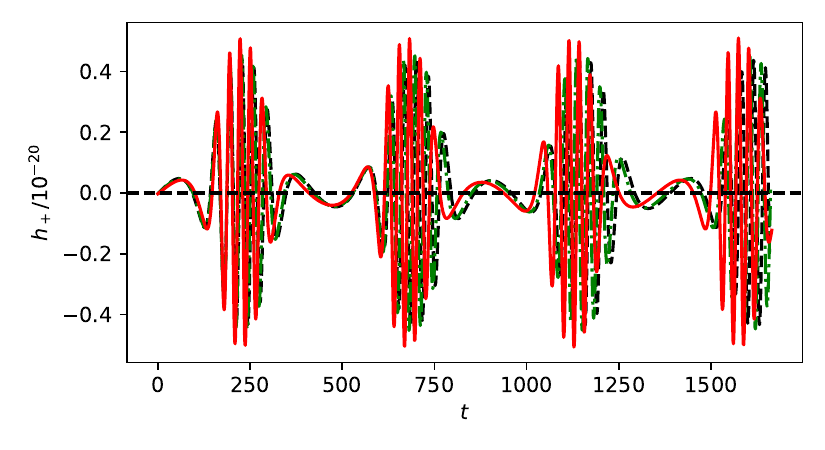}\\ 
        \includegraphics[width=0.45\textwidth]{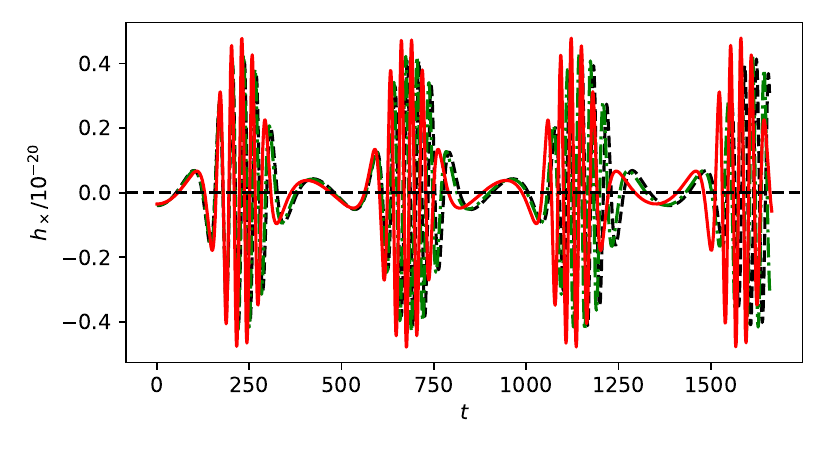} 
    \end{tabular}
    \caption{The figure illustrates the effect of the charge parameter, or the effect of the polymerization parameter $\delta_b$, on the periodic orbit $(3, 2, 1)$. The right panel shows the periodic orbit and the left panel displays the corresponding GW polarizations $h_+$ and $h_\times$ in the time--domain. The parameters are taken as $E = 0.96$, $m = 10 M_\odot$, $M = 10^6 M_\odot$, and different values of charge as $Q = 0.0$, $0.3$, and $0.6$. By increasing $Q$, the amplitude of the waveform enhances and a phase shift is observed.} 
    \label{fig:orbitGW_Q}
\end{figure*}

\begin{itemize}
    \item \textbf{Amplitude enhancement:} As $Q$ increases, the peak amplitudes of $h_+$ and $h_\times$ during the whirl phase get larger. From the minimal area condition~\eqref{eq:deltab}, a bigger $Q$ at fixed $M$ means a bigger polymerization parameter $\delta_b$ (and thus $b_0$). That changes the quantum correction factor $[1 - (r_0/M)g(r)]$ in the metric. It also shifts the periapsis radius and changes how fast the particle moves at closest approach. The result is stronger gravitational wave emission during the whirl phase. So the amplitude boost comes from LQG polymerization, not just from classical charge effects~\cite{Yang:2024cnd, Ahmed:2025shr}.

    \item \textbf{Phase shift:} The waveforms also show a phase shift that grows with $Q$ (and therefore $\delta_b$). This comes from the quantum--corrected metric modifying the orbital frequencies $\omega_r$ and $\omega_\phi$. Even though the frequency shift per cycle is tiny, it adds up coherently over the many cycles of an EMRI inspiral~\cite{Barack:2003fp, Babak:2006uv, Yang:2024cnd}. And since $\delta_b$ grows monotonically with $Q$ through Eq.~\eqref{eq:deltab}, the resulting phase drift could be a useful way to probe the LQG parameter space.
\end{itemize}

\subsection{Frequency--domain analysis and detector sensitivity}

\begin{figure}
    \centering
    \subfigure{\includegraphics[width=0.9\linewidth]{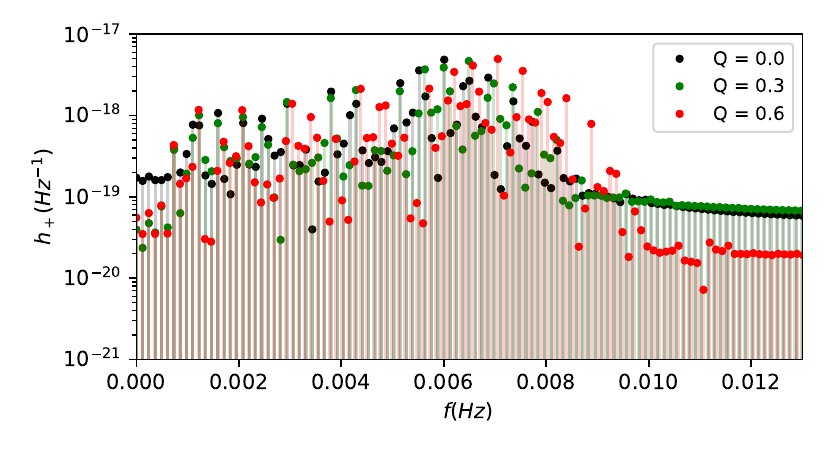}}
    \subfigure{\includegraphics[width=0.9\linewidth]{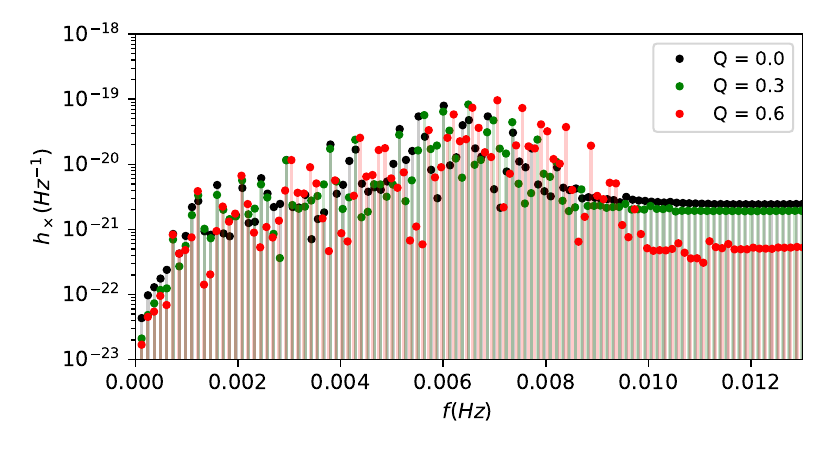}}
    \caption{The amplitude spectra $|h_+|$ and $|h_\times|$ are obtained in the frequency--domain by applying discrete Fourier transform on the GW polarizations displayed in Fig.\ref{fig:orbitGW_Q}. The figure shows the effect of the charge $Q$ parameter on the spectral content.}
    \label{fig:hphc_f}
\end{figure}

To see whether our predicted signals could actually be detected, we convert the time--domain waveforms to the frequency--domain using a discrete Fourier transform. This gives us the spectral amplitudes $|\tilde{h}_+(f)|$ and $|\tilde{h}_\times(f)|$. From these, we compute the characteristic strain $h_c(f)$, the standard way to compare a source signal to detector noise~\cite{Maggiore:2007ulw, Thorne:1997ut}:
\begin{equation}\label{eq:characteristic_strain}
    h_c(f) = 2f\sqrt{|\tilde{h}_+(f)|^2 + |\tilde{h}_\times(f)|^2}.
\end{equation}
Figure~\ref{fig:hphc_f} shows the frequency--domain spectra for the $(3,2,2)$ orbit at different $Q$ values. In all cases, the main spectral peak falls in the millihertz band. 
This is consistent with the expected frequency range of EMRIs around supermassive black holes, whose characteristic frequencies scale as $f \sim (M/10^6\,M_\odot)^{-1}\,\mathrm{mHz}$. This puts the signal right in the sensitivity windows of LISA, Taiji, and TianQin~\cite{LISA:2017pwj, Hu:2017mde, TianQin:2015yph, LISA:2022yao}.

\begin{figure}[ht]
    \centering
    \includegraphics[width=0.9\linewidth]{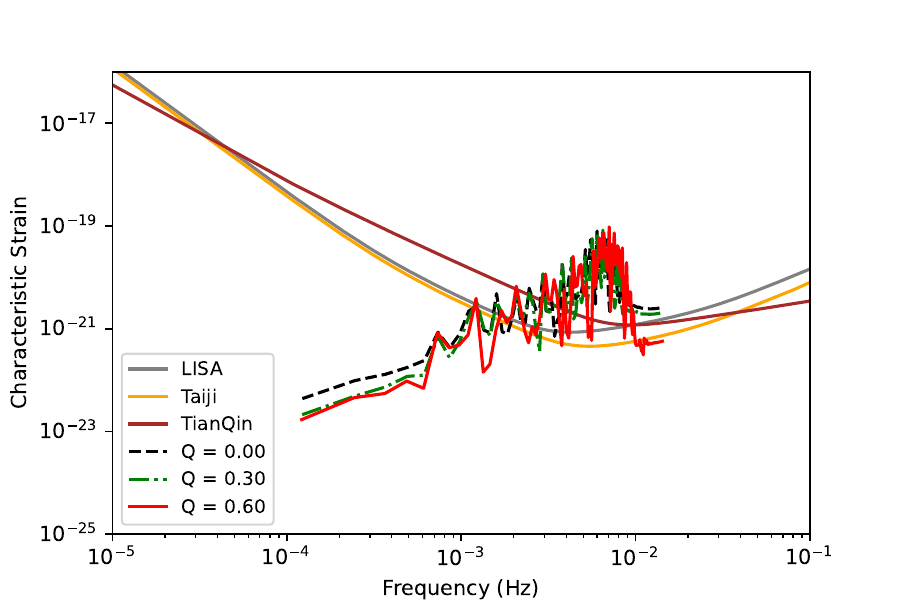}
    \caption{The characteristic strain for the $(3,2,1)$ periodic orbit is plotted versus the frequency for different values of the charge parameter as $Q = 0.0$ (black dashed line), $Q = 0.3$ (green dot-dashed line), and $Q = 0.6$ (red solid line). The resulting strains have a peak at the millihertz range and they are crossing the sensitivity curves of the future space-based GW detectors LISA, Taiji, and TianQin. The peak enhances with increasing the charge parameter; in other words by increasing the polymerization parameter. }
    \label{fig:strain}
\end{figure}

\begin{figure}[ht]
    \centering
    \includegraphics[width=0.9\linewidth]{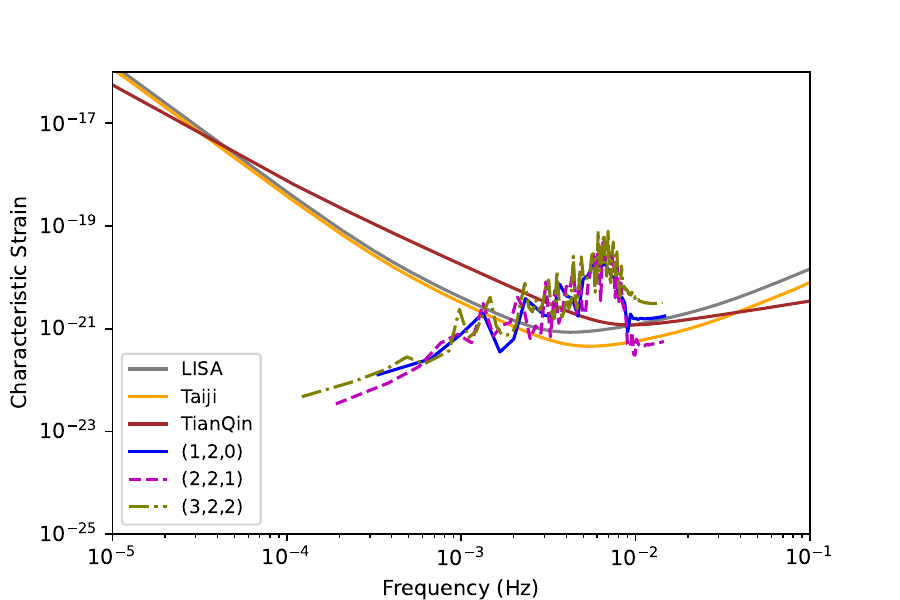}
    \caption{The characteristic strain is displayed for three periodic orbits $(z,w,v) = (1,2,0)$, $(2,2,1)$, and $(3,2,2)$ where $E = 0.96$ and $Q = 0.4$. By increasing the zoom number, the orbit gets more complex and the resulting characteristic strain enhances.  }
    \label{fig:stain_zwv}
\end{figure}

The corresponding characteristic strain comes from the frequency--domain polarizations $\tilde{h}_+(f)$ and $\tilde{h}_\times(f)$ we just computed. Figure~\ref{fig:strain} plots the results. The peak of the signal lines up with the projected noise curves of all three detectors. The characteristic strain for three different periodic orbits $(z,w,v) = (1,2,0)$, $(2,2,1)$, and $(3,2,2)$  are illustrated in Fig. \ref{fig:stain_zwv}. The energy and charge are respectively taken as $E = 0.96$ and $Q = 0.4$. The characteristic strain is obtained by applying the discrete Fourier transform on the corresponding gravitational wave polarizations, as shown in Fig. \ref{fig:hchp_zwv}. It is realized that by increasing the zoom number, the orbit gets more complex and the resulting characteristic strain enhances. A few things stand out.

\begin{itemize}
    \item \textbf{Detectability:} For every orbit family and parameter value, we looked at, the peak characteristic strain sits above the sensitivity thresholds of LISA, Taiji, and TianQin. That means gravitational waves from periodic orbits in this charged LQG spacetime should in principle be detectable by future space--based missions~\cite{LISA:2017pwj, Hu:2017mde, TianQin:2015yph, LISA:2022yao}.
    
    \item \textbf{Charge parameter dependence:} Both the peak amplitude and the peak frequency of $h_c$ go up as $Q$ increases. Larger $Q$ gives a higher, slightly blue--shifted characteristic strain peak. This systematic trend means that precise measurements of the peak amplitude and frequency could help constrain the polymerization parameter, and by extension, the LQG minimal area scale~\cite{Yang:2024cnd, Chen:2026kbn}.
    
    \item \textbf{Orbit topology dependence:} For fixed LQG parameters, the characteristic strain changes a lot between different periodic orbit families (Fig.~\ref{fig:stain_zwv}). So the zoom--whirl structure of the orbit directly shapes the spectral content and amplitude of the detectable gravitational wave signal. Different orbit topologies leave different fingerprints in the frequency--domain~\cite{Yang:2024cnd, Ahmed:2025shr}.
\end{itemize}

Taken together, these results show that gravitational waves from periodic orbits in the charged LQG spacetime carry clear imprints of quantum corrections, including amplitude enhancement, phase drift, and spectral shifts, all of which differ systematically from classical GR predictions. With the sensitivity improvements expected from future space--based detectors, EMRI observations could offer a concrete way to probe the LQG polymerization parameter in the strong--field regime~\cite{Yang:2024cnd, Chen:2026kbn, LISA:2022yao}.

\section{Conclusion}
\label{sec:conclusion}
GR works spectacularly well in weak fields, but we know it can't be the full story. Singularities signal its breakdown, and quantum mechanics has to enter somewhere. LQG is a serious candidate for a quantum theory of gravity, but like any such theory, it needs observational constraints. 
Black holes provide a natural laboratory for probing quantum-gravity effects, because their strong gravitational fields amplify Planck-scale effects. EMRIs are particularly promising because they spend a long time in the strong field, accumulating small deviations into something measurable. Our goal was to see whether the polymerization scheme in LQG leaves an imprint on EMRI gravitational waves that future detectors could actually detect.

Motivated by this, we determine whether gravitational waves from EMRI could reveal something about LQG. If LQG modifies the spacetime geometry around a black hole, those modifications should leave a mark on the gravitational waves emitted by a smaller body orbiting it. 
We therefore consider the LQG-inspired charged black hole constructed by Borges \emph{et al.}~\cite{Borges:2023fub}, obtained by applying polymerization techniques to the Reissner–Nordström Hamiltonian, and worked out the details.

The polymerization parameter $\delta_b$ is not free here. The minimal area condition ties it directly to the black hole parameters mass $M$ and charge $Q$. 
So, by varying $Q$, we are not only changing the electric charge, but also modifying the strength of the LQG corrections. Even for $Q=0$, the corrections do not disappear, which can be confirmed from the Table. \ref{tab:params}. That's important because it means the quantum effects are baked into the geometry from the start, not just by some arbitrary parameter.

We started by looking at timelike geodesics in LQG spacetime and found that the quantum correction factor $[1 - (r_0/M)g(r)]$ appears directly in the effective potential, changing everything that follows. We computed the ISCO and MBO radii, energies, and angular momenta for different $Q$. As $Q$ goes up, both $r_{\rm ISCO}$ and $r_{\rm MBO}$ shrink. The allowed $(E,L)$ region for bound orbits also gets wider. So LQG does two things: it pulls the strong-field region inward, and it changes the orbital parameters which actually matter.

The periodic orbits are classified using the Levin-Perez-Giz $(z,w,v)$ taxonomy. The frequency ratio $q$ grows with energy and drops with angular momentum; nothing surprising there. But the maximum allowed energy for a bound orbit decreases when $Q$ goes up, which is a direct consequence of the LQG corrections. We plotted a bunch of periodic orbits for different $(z,w,v)$ combinations. The patterns get more complicated as $z$ increases, and larger $w$ adds more whirl loops near the black hole. All of this is standard for strong-field relativity, but the LQG corrections shift the specific energies and angular momenta at which these orbits occur.

We used the numerical kludge method, integrate the geodesics, and then substitute into the quadrupole formula. The waveforms have that classic zoom-whirl look. Most of the time, low smooth oscillations occur while the particle is far out. Then, wham, and a  sharp burst when it swoops close to the black hole. Number of bursts per cycle = $w$, exactly what the taxonomy says.

The LQG corrections show up in two clear ways. First, the burst amplitudes increase with $Q$. This is not just a charge effect. It comes from the quantum correction factor in the metric shifting the periapsis radius and speeding up the particle at closest approach. Second, the waveforms develop a phase shift that also grows with $Q$. The frequency shift per orbit is tiny, but it adds up over the many cycles of an EMRI inspiral. Both effects are small individually, but coherent accumulation makes them potentially measurable.

We also looked at the frequency-domain signals. The characteristic strain peaks in the millihertz band for all cases; right where LISA, Taiji, and TianQin are most sensitive. In fact, the peak strain sits above the noise curves for all three detectors. That means if an EMRI like this exists out there, these detectors should see it. We also noticed that both the peak amplitude and the peak frequency increase with $Q$, and that the strain spectrum changes noticeably across different orbit topologies. So the signal is not just a yes/no detection, but it carries detailed information about the underlying spacetime geometry.

So where does this leave us? We've shown that periodic orbits in a charged LQG black hole produce GW signatures that differ systematically from classical GR. The differences come in the form of amplitude boosts, phase drifts, and spectral shifts. They are small, but they add up over time, and future space--based detectors will have the sensitivity to detect them. That opens up a real possibility: using EMRI observations to constrain the LQG polymerization parameter. The exact numbers will depend on what the detectors actually see, but the fact that the signal exceeds the sensitivity bounds for LISA, Taiji, and TianQin means the constraints could be meaningful.

It is encouraging that these effects emerge at this level. Orbital frequency shifts change the phase. The modified geometry changes the amplitude and spectrum. And it all gets bigger when $Q$ (and $\delta_b$) goes up. Will it survive in more realistic calculations? No idea. But this shows quantum gravity can leave a mark on something EMRI measurements might actually catch.

Looking ahead, there's plenty more to do. The natural next step is to include radiation reaction and see how the orbit evolves over the full inspiral, not just a single geodesic. That would tell us whether the phase shifts we observed in individual orbits persist as the orbit shrinks. Also, we only looked at the charged version of this LQG black hole. The neutral case already shows quantum corrections, so a systematic comparison between charged and neutral models would help isolate the LQG effects from classical electromagnetic effects. And finally, a full parameter estimation study, including injecting fake signals into simulated LISA data and seeing how well we can recover $\delta_b$, would give us a real sense of the constraints we could actually place.

Periodic orbits in LQG black holes leave a mark we can actually observe. The signals are strong enough. The quantum imprints are clear enough. And the detectors are almost here. Maybe in a few years, we will be using EMRIs to poke at quantum gravity; not just in theory, but in real data.

\section*{Acknowledgements}
SGG would like to thank the Institute for Theoretical Physics and Cosmology, Zhejiang University of Technology, Hangzhou 310023, China, for its hospitality during the collaboration visit, during which part of this work was conducted.

\bibliographystyle{apsrev4-1}
\bibliography{main_bib.bib}

\end{document}